\documentclass[aps,footinbib,floatfix,twocolumn]{revtex4}
\usepackage{amssymb}
\usepackage[pdftex]{graphicx}	
	\DeclareGraphicsExtensions{.pdf,.png,.jpg}
\usepackage[applemac]{inputenc}
\usepackage{amsmath,amssymb,amsthm}
\usepackage{mathbbol,bbm}
\usepackage{graphicx,float}
\usepackage[final]{showkeys}
\usepackage{bm,dsfont}
\usepackage{color}
\usepackage{hyperref}
\usepackage{makecell}
\usepackage{enumitem}
\usepackage{listings}

\usepackage{color}
\definecolor{myblue}{rgb}{0.3,0.4,0.85}
\definecolor{myred}{rgb}{0.8,0.,0.2}
\definecolor{mygreen}{rgb}{0.6,0.8,0.2}

\def\ket#1{\left|#1\right>}

\def\Tr{ {\rm{Tr }}}

\usepackage{booktabs}

\begin{document}
\title{Observation of topological Uhlmann phases with superconducting qubits}
\author{O. Viyuela$^{1,4,5}$\footnote{Correspondence to oviyuela@mit.edu}, A. Rivas$^1$, S. Gasparinetti$^2$, A. Wallraff$^2$, S. Filipp$^3$ and M.A. Martin-Delgado$^1$}
\affiliation{1. Departamento de F\'{\i}sica Te\'orica I, Universidad Complutense, 28040 Madrid, Spain\\
2. Department of Physics, ETH Zurich, CH-8093 Zurich, Switzerland\\
3. IBM Research - Zurich, 8803 Rueschlikon, Switzerland\\
4. Department of Physics, Massachusetts Institute of Technology, Cambridge, MA 02139, USA\\
5. Department of Physics, Harvard University, Cambridge, MA 02318, USA}

\vspace{-3.5cm}

\begin{abstract}
Topological insulators and superconductors at finite temperature can be characterised by the topological Uhlmann phase. However, a direct experimental measurement of this invariant has remained elusive in condensed matter systems. Here, we report a measurement of the topological Uhlmann phase for a topological insulator simulated by a system of entangled qubits in the \emph{IBM Quantum Experience} platform. By making use of ancilla states, otherwise unobservable phases carrying topological information about the system become accessible, enabling the experimental determination of a complete phase diagram including environmental effects. We employ a state-independent measurement protocol which does not involve prior knowledge of the system state. The proposed measurement scheme is extensible to interacting particles and topological models with a large number of bands. 
\end{abstract}

\maketitle


\section*{INTRODUCTION}

The search for topological phases in condensed matter \cite{RMP,Bernevig_et_al06,Koenig_et_al07,Hsieh_et_al12,Dziawa_et_al12,Xu_et_al15} has triggered an experimental race to detect and measure topological phenomena in a wide variety of quantum simulation experiments \cite{Atala_et_al_12,Jotzu_et_al_14,Duca_et_al_15,Schroer_et_al14,Roushan_et_al14,Li_et_al_16,Flurin_et_al_17}.
In quantum simulators the phase of the wave function can be accessed directly, opening a whole new way to observe topological properties \cite{Leek_et_al07,Atala_et_al_12,Duca_et_al_15} beyond the realm of traditional condensed matter scenarios. These quantum phases are very fragile, but when controlled and mastered, they can produce very powerful computational systems like a quantum computer \cite{NC,rmp_GMA}. The Berry phase \cite{Berry84} is a special instance of quantum phase, that is purely geometrical \cite{WilczekBook} and independent of dynamical contributions during the time evolution of a quantum system. In addition, if that phase is invariant under deformations of the path traced out by the system during its evolution, it becomes topological. Topological Berry phases have also acquired a great relevance in condensed matter systems. The now very active field of topological insulators (TIs) and superconductors (TSCs) \cite{RMP} ultimately owes its topological character to Berry phases \cite{Zak89} associated to the special band structure of these exotic materials.

However, if the interaction of a TI or a TSC with its environment is not negligible, the effect of the external noise in the form of e.g. thermal fluctuations, makes these quantum phases very fragile \cite{Viyuela_et_al12,Rivas_et_al13,Mazza_et_al13,Bardyn_et_al13,Evert_et_al14,Shen_et_al14,Dehghani_et_al14,Hu_et_al15,Victor_et_al16,Linzner_et_al_16,Claeys_et_al16,Lemini_et_al_16,Bardyn_et_al17}, and they may not even be well-defined. For the Berry phase acquired by a pure state, this problem has been successfully adressed for one-dimensional systems \cite{Viyuela_et_al14} and extended to two-dimensions later  \cite{Arovas14,Viyuela_et_al14_2D,Viyuela_et_al15}. The key concept behind this theoretical characterisation is the notion of \emph{Uhlmann phase} \cite{Uhlmann86,Soqvist_et_al_00,Ericsson2003,Aberg_et_al07,Zhu_et_al11,Budich_et_al15,Ericsson_et_al16,Mera_et_al16,Mera_et_al17}, a natural extension of the Berry phase for density matrices. In analogy to the Berry phase, when the Uhlmann phase for mixed states remains invariant under deformations, it becomes topological.

Although this phase is gauge invariant and thus, in principle, observable, a fundamental question remains: how to measure a topological Uhlmann phase in a physical system? To this end, we employ an ancillary system as a part of the measurement apparatus. By encoding the temperature (or mixedness) of the system in the entanglement with the ancilla, we find that the Uhlmann phase appears as a relative phase that can be retrieved by interferometric techniques. The difficulty with this type of measurement is that it requires a high level of control over the environmental degrees of freedom, beyond the reach of condensed matter experiments. On the contrary, this situation is especially well-suited for a quantum simulation scenario. 

Specifically, in this work we report: i) the measurement of the topological Uhlmann phase on a quantum simulator based on superconducting qubits \cite{Schoelkopf_et_al08,Barends_et_al15,Salathe_et_al15}, in which we have direct control over both system and ancilla, and ii) the computation of the topological phase diagram for qubits with an arbitrary noise degree. A summary and a comparison with pure state topological measures are shown in Fig ~\ref{Fig_diagram}. 
In addition, we construct a state independent protocol that detects whether a given mixed state is topological in the Uhlmann sense.
Our proposal also provides a quantum simulation of the AIII class \cite{Ludwig,Kitaev_2009} of topological insulators (those with chiral symmetry) in the presence of disturbing external noise. Other cases of two-dimensional TIs, TSCs and interacting systems can also be addressed by appropriate modifications as mentioned in the conclusions.

\section*{RESULTS}

\subsection*{Topological Uhlmann phase for qubits}

We briefly present the main ideas of the Uhlmann approach for a two-band model of TIs and TSCs simulated with a qubit. Let $\theta(t)|_{t=0}^1$ define a closed trajectory along a family of single qubit density matrices parametrised by $\theta$,  
\begin{equation}\label{rho1}
\rho_\theta=(1-r)|0_{\theta}\rangle\langle 0_{\theta}| + r|1_{\theta}\rangle\langle 1_{\theta}|,
\end{equation}
where $r$ stands for the mixedness parameter between the $\theta$-dependent eigenstates  $|1_{\theta}\rangle$ and $|0_{\theta}\rangle$, e.g. that of a transmon qubit \cite{Koch_et_al07}. The mixed state $\rho_{\theta}$ can be seen as a ``part'' of a state vector $|\Psi_{\theta}\rangle$ in an enlarged Hilbert space ${\cal H}={\cal H}_{\rm S}\otimes{\cal H}_{\rm A}$, where S stands for system and A for the ancilla degrees of freedom with $\dim{\cal H}_{\rm A}\geq\dim{\cal H}_{\rm S}$. The state vector $|\Psi_{\theta}\rangle$ is a so-called purification of $\rho_{\theta}={\rm Tr_A}\Big(|\Psi_{\theta}\rangle\langle\Psi_{\theta}|\Big)$, where ${\rm Tr_A}$ performs the partial trace over the ancilla. There is an infinite number of purifications for every single density matrix, specifically $(\mathds{1}\otimes{U}_{\rm A})|\Psi_{\theta}\rangle$ for any unitary $U_{\rm A}$ acting on the ancilla purifies the same mixed state as $|\Psi_{\theta}\rangle$. Hence, for a family of density matrices $\rho_{\theta}$, there are several sets of purifications $|\Psi_{\theta}\rangle$ according to a U(n) gauge freedom. This generalizes the standard U(1) gauge (phase) freedom of state vectors describing quantum pure states to the general case of density matrices.

Along a trajectory $\theta(t)|_{t=0}^1$ for $\rho_\theta$ the induced purification evolution (system qubit S and ancilla qubit A) can be written as
\begin{equation}
\begin{split}
|\Psi_{\theta(t)}\rangle&=\sqrt{1-r}U_{\rm S}(t)\ket{0}_{\rm S}\otimes U_{\rm A}(t)\ket{0}_{\rm A} +\\
&+ \sqrt{r}U_{\rm S}(t)\ket{1}_{\rm S}\otimes U_{\rm A}(t)\ket{1}_{\rm A},
\end{split}
\label{psi2}
\end{equation}
where $\ket{0}=\begin{pmatrix}1 \\ 0 \end{pmatrix}$ and $\ket{1}=\begin{pmatrix}0 \\ 1 \end{pmatrix}$ is the standard qubit basis, and $U_{\rm S}(t)$ is a unitary matrix determined by the $\theta$-dependence. Moreover the arbitrary unitaries $U_{\rm A}(t)$ can be selected to fulfill the so-called Uhlmann parallel transport condition. Namely, analogously to the standard Berry case, the Uhlmann parallel transport requires that the distance between two infinitesimally close purifications $\| |\Psi_{\theta(t+dt)}\rangle-|\Psi_{\theta(t)}\rangle \|^2$ reaches a minimum value (which leads to removing the relative infinitesimal ``phase'' between purifications) \cite{Uhlmann86}. Physically, this condition ensures that the accumulated quantum phase (the so-called Uhlmann phase $\Phi_{\rm U}$) along the trajectory is purely geometrical, that is, without dynamical contributions. This is a source of robustness, since variations on the transport velocity will not change the resulting phase. 

\begin{figure}[t]
\includegraphics[width=0.60\columnwidth]{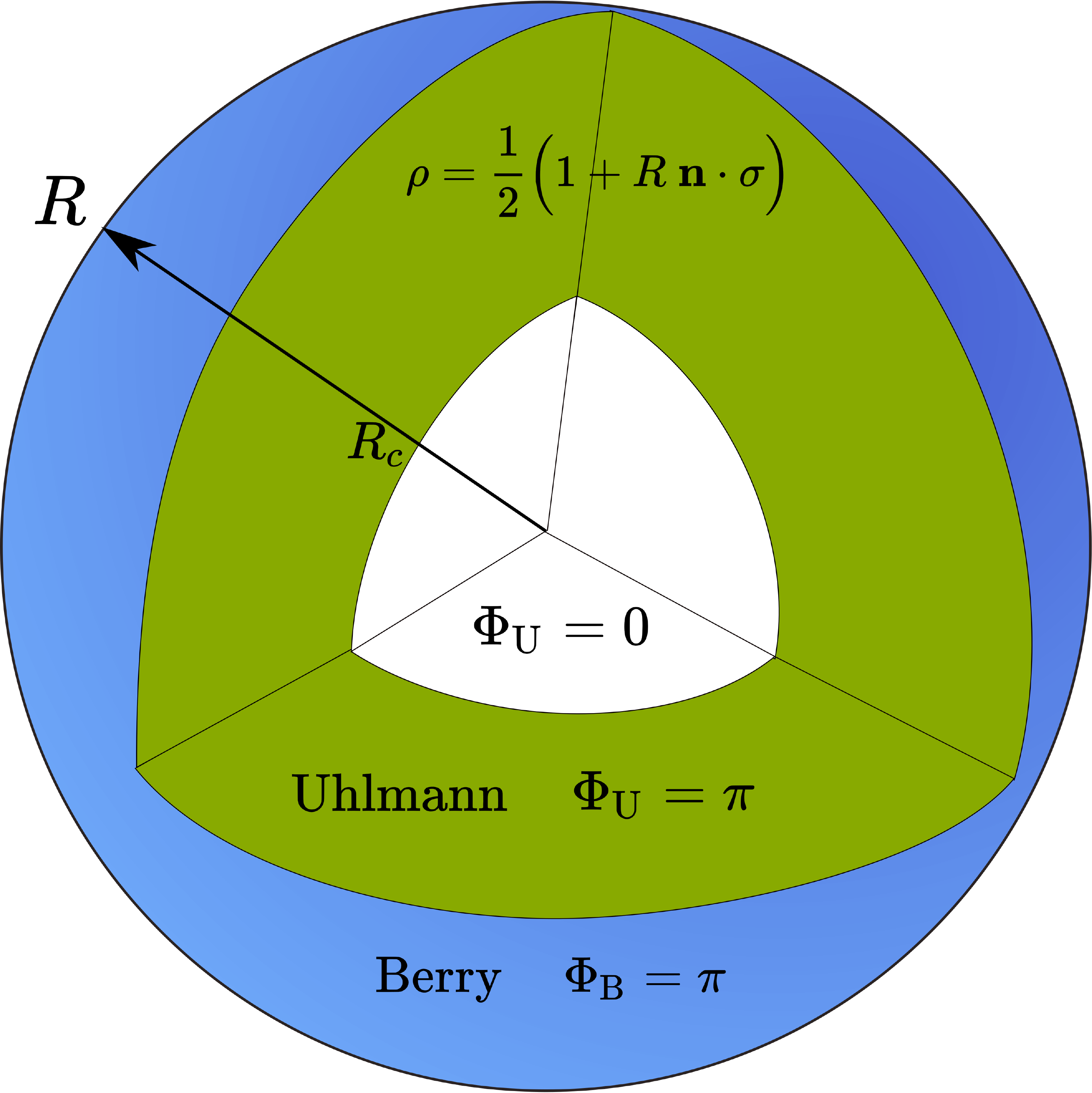}
\caption{Topological measures for a single qubit in a mixed state $\rho=(1-r)|1\rangle\langle 1|+r|0\rangle\langle 0|=\frac{1}{2}\Big({\mathbb 1} + R~{\bf n}\cdot{\bf \sigma}\Big)$ in the Bloch sphere representation. The mixedness parameter $r$ between states $|1\rangle$ and $|0\rangle$ is encoded into the degree of mixedness $R=|2r-1|$ . We compute the Berry $\Phi_{\rm B}$ and Uhlmann $\Phi_{\rm U}$ phases for non-trivial topological regimes. If $r\not\in\{1,0\}$ or equivalently $R<1$, then only $\Phi_{\rm U}$ is well defined and highlights a non-trivial topological phase ($\Phi_{\rm U}=\pi$), provided that $R>R_c$. Here, $R_c$ denotes the critical amount of noise that the system can withstand while remaining topological.}
\label{Fig_diagram}
\end{figure}

Next, we consider the Hamiltonian of a two-band topological insulator in the AIII chiral-unitary class \cite{Ludwig,Kitaev_2009}, $H=\sum_k\Psi_k^{\dagger}H_k\Psi_k$, in the spinor representation $\Psi_k=(\hat{a}_k,\hat{b}_k)^{\rm t}$ where $\hat{a}_k$ and $\hat{b}_k$ stands for two species of fermionic operators. The one-particle Hamiltonian is 
\begin{align} \label{H1}
H_k&=\frac{G_k}{2}{\bm n}_k\cdot{\bm \sigma},\nonumber\\
{\bm n}_k&=\frac{2}{G_k}(\sin{k},0,M+\cos{k}),\\G_k&=2\sqrt{1 + M^2 + 2M\cos{k}}. \nonumber
\end{align}
where $G_k$ represents the actual gap between the valence and conduction bands in the topological insulator, and ${\bm n}_k$ is a unit vector called winding vector \cite{Viyuela_et_al14}. We now map the crystalline momentum $k$ of the topological insulator \cite{RMP} to a tunable time-dependent paramenter $\theta$ of the quantum simulator. When invoking the rotating wave approximation this model also describes, e.g. the dynamics of a driven transmon qubit \cite{Schroer_et_al14,Leek_et_al07}. The detuning $\Delta=2\big(\cos{\theta} + M\big)$ between qubit and drive is parametrised in terms of $\theta$ and a hopping amplitude $M$, whereas the coupling strength between the qubit and the incident microwave field is given by $\Omega=2\sin{\theta}$.

The non-trivial topology of pure quantum states ($r\in\{0,1\}$) of this class of topological materials can be witnessed by the winding number. This is defined as the angle swept out by ${\bm n}_\theta$ as $\theta$ varies from $0$ to $2\pi$, namely,
\begin{equation}
\omega_1:=\frac{1}{2\pi}\oint\bigg(\frac{\partial_\theta {n}_\theta^x}{n_\theta^z}\bigg)d\theta.
\label{w1}
\end{equation}
Then, using Eq.~\eqref{H1} and Eq.~\eqref{w1}, the system is topological ($\omega_1=1$) when the hopping amplitude is less than unity ($M<1$) and trivial ($\omega_1=0$) if $M>1$. In fact, the topological phase diagram coincides with the one given by the Berry phase acquired by the ``ground'' state $|0\rangle_{\theta}$ (or the ``excited'' state $|1\rangle_{\theta}$) of Hamiltonian (\ref{H1}) when $\theta$ varies from $0$ to $2\pi$, (see Supplementary Note 2).

The computation of the unitary $U_{\rm S}$ in Eq. \eqref{psi2} for a transportation in time of $\theta$ according to the Hamiltonian \eqref{H1} yields
\begin{equation}
U_{\rm S}(t)={\rm e}^{-{\rm i}\int_0^{t}h(t')dt'{\bm \sigma_y}},
\label{Uas}
\end{equation}
with $h(t):=\frac{\partial_{t} {n}_{t}^x}{2n_{t}^z}$. This implements the eigenstate transport $|1_{\theta(t)}\rangle=U_{\rm S}(t)|1\rangle$ and $|0_{\theta(t)}\rangle=U_{\rm S}(t)|0\rangle$. In addition, we can consider a similar form for the unitary $U_{\rm A}$ in Eq. \eqref{psi2},
\begin{equation}
U_{\rm A}(t)=[U_{\rm S}(t)]^{p_a}={\rm e}^{-{\rm i}\int_0^{t}p_a h(t')dt'{\bm \sigma_y}},
\label{Uas2}
\end{equation}
where the parameter $p_a \in[0,1]$ is defined as an ancillary ``weight''. We find that the Uhlmann parallel transport condition is satisfied for $p_a=p_r:=2\sqrt{r(1-r)}$. The detailed technical derivation is provided in Supplementary Notes 1 and 2. 

Now, from Eq.~\eqref{psi2} it is possible to define the relative phase $\Phi_{\rm M}$ between the initial $|\Psi_{\theta(0)}\rangle$ and the final state, i.e. $|\Psi_{\theta(t_{\rm f})}\rangle$. For Hamiltonian \eqref{H1}, density matrix \eqref{rho1} and purification \eqref{psi2}, we find
\begin{eqnarray}
\label{phase_overlap2}
\Phi_{\rm M}&:=&\arg{[\langle\Psi_{\theta(0)}|\Psi_{\theta(t_{\rm f})}\rangle]}=\\
&=&\arg{\Big[\cos{(I_{0}^{t_{\rm f}})}\cos{(p_aI_{0}^{t_{\rm f}})}+p_r\sin{(I_{0}^{t_{\rm f}})}\sin{(p_aI_{0}^{t_{\rm f}})}\Big]},
\nonumber
\end{eqnarray}
where $I_{t_0}^{t_{\rm f}}:=\int_{t_0}^{t_{\rm f}}h(t')dt'$. As commented before, by assuming $p_a=p_r:=2\sqrt{r(1-r)}$, the purification precisely follows Uhlmann parallel transport and the relative phase $\Phi_{\rm M}$ becomes the Uhlmann phase $\Phi_{\rm U}$ associated to the trajectory. For a closed path $t_{\rm f}=1$, the integral $I_{0}^1=\pi\omega_1=\Phi_{B}$ becomes the topological Berry phase. In that case, the Uhlmann phase simplifies to
\begin{equation}
\Phi_{\rm U}=\arg\{\cos[(1-2p_r)\pi \omega_1]\}.
\label{phiU3}
\end{equation}
We can now deduce the topological properties of these phases in the presence of external noise, as measured by the parameter $r$ [Eq.~\eqref{rho1}]. This is depicted in Fig.~\ref{Fig_diagram}.
Namely, if $M>1$ then $\omega_1=0$, and $\Phi_{\rm U}=0$ (trivial phase) for every mixedness parameter $r$. If $M<1$ then $\omega_1=1$ and one obtains $\Phi_{\rm U}=\arg[-\cos(2\pi \sqrt{r(1-r)})]$. If the state is pure ($r=0$), then $\Phi_{\rm U}^{0}=\pi$, recovering the same topological phase given by the winding number and the Berry phase. However, for $r\not= 0$ there are critical values of the mixedness $r_c$ at which the Uhlmann phase, according to Eq.~\eqref{phiU3}, jumps from $\pi$ to zero (see Fig.~\ref{Fig_diagram}). The first $r_{c1}=\frac{1}{4}(2-\sqrt{3})\approx0.067$ signals the mixedness at which the system loses the topological character of the ground state. Moreover, there exists another $r_{c2}=1-r_{c1}$ at which the system becomes topological again due to the topological character of the excited state ($r\rightarrow1$). Notice that at $r=1$ the system becomes a pure state again (the excited state), which is also topologically non-trivial according to the Berry phase. Actually, provided that the weight $p_r<p_{r=r_{c1(2)}}=0.5$, the system is topological in the Uhlmann sense as long as $M<1$.
This reentrance in the topological phase at $r_{c2}$ was absent in previous works \cite{Viyuela_et_al14,Viyuela_et_al14_2D,Arovas14}.

\begin{figure}[t]
\includegraphics[width=\columnwidth]{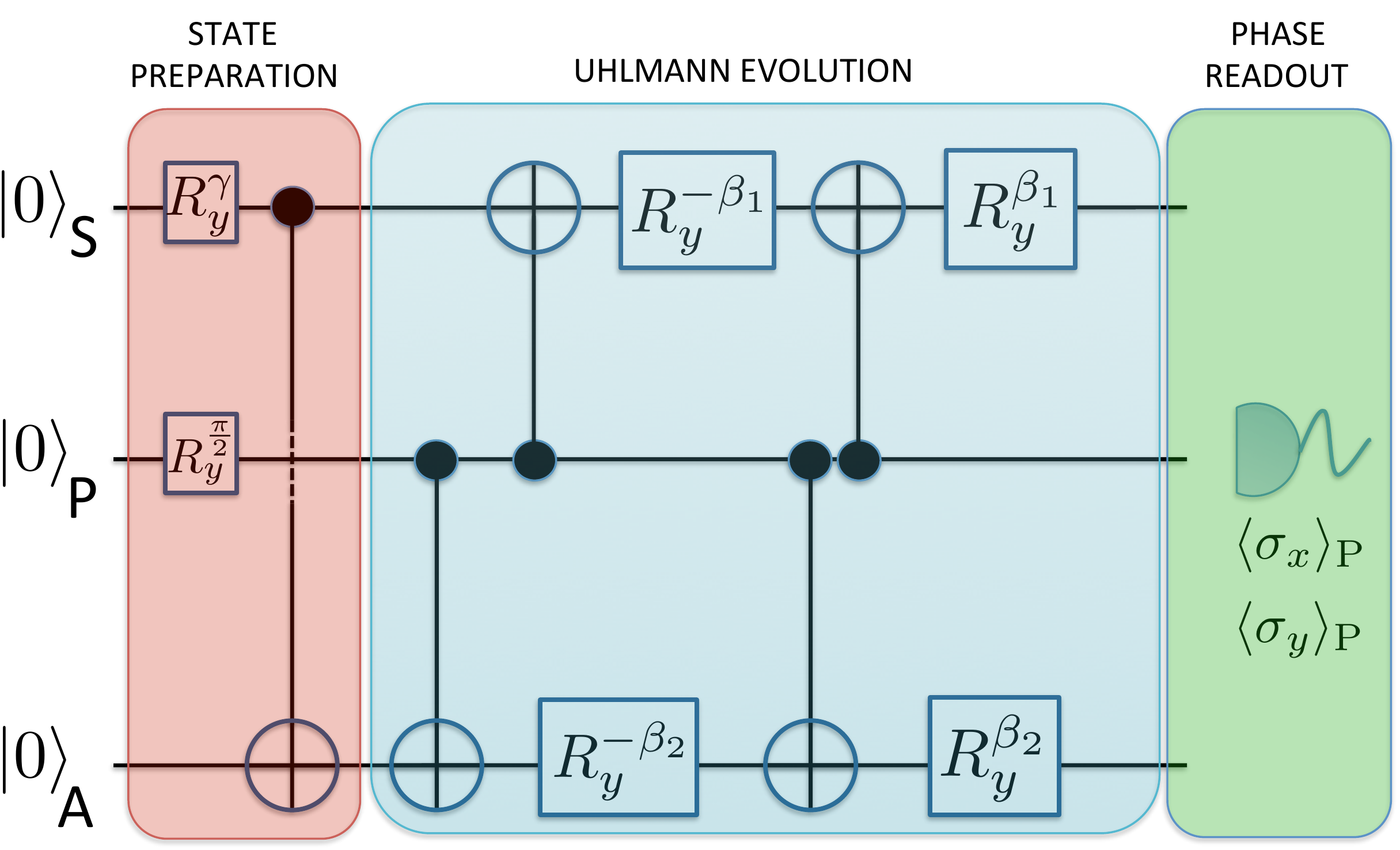}
\caption{Circuit diagram to measure the topological Uhlmann phase, e.g., with superconducting circuits as explained in the text. The circuit represents the decomposition of the bi-local unitary evolution $U_{\rm S}(t)\otimes U_{\rm A}(t)$, defined in Eq.~\eqref{Uas}, into elementary single and two-qubit CNOT gates \cite{rmp_GMA}. The gate $R_y^\gamma$ represents a single qubit rotation about the y-axis for an angle $\gamma=2~\text{arccos}{\sqrt{1-r}}$, and the angles $\beta_1$ and $\beta_2$ appear in Eq.~\eqref{phase_overlap2}.} 
\label{Fig_circuit}
\end{figure}

\subsection*{Experimental realization}

Measuring the topological Uhlmann phase is a very challenging task since its definition in terms of purifications implies precise control over auxiliary/environmental degrees of freedom (the ancilla). In an experiment, we therefore include an extra ancilla qubit representing the environment. We also include a third qubit acting as a probe system $P$, such that by measuring qubit $P$ we retrieve the accumulated phase by means of interferometric techniques. The measurement protocol is described in Fig.~\ref{Fig_circuit}:

\textbf{Step 1.} Following Eq.~\eqref{psi2}, we prepare the initial state $|\Psi_{\theta(0)}\rangle \otimes \ket{0}_{\rm P}$ (red block of Fig.~\ref{Fig_circuit}) using single qubit rotations $R_y^\gamma$ about the y-axis for an angle $\gamma=2\text{arcos}{\sqrt{1-r}}$ and a two-qubit controlled not gate. For superconducting qubits, the latter can be performed e.g. by implementing a controlled phase gate for frequency-tunable transmons \cite{Strauch_et_al07} or by a cross-resonance gate \cite{Chow_et_al11}.

\textbf{Step 2.} We apply the bi-local unitary $U_{\rm S}(t)\otimes U_{\rm A}(t)$ on $S\otimes A$ conditional to the state of the probe $P$. This is accomplished by single qubit rotations about an angle $\beta_1$ or $\beta_2$, determined by $h(t)$ and $p_a$ (blue block of Fig.~\ref{Fig_circuit}), and two-qubit gates. This decomposition is based on the fact that any controlled unitary gate can be always decomposed as a product of unitary single-qubit gates and two-qubit CNOT gates \cite{rmp_GMA}. Fig.~\ref{Fig_circuit} shows the final result after the decomposition of the Uhlmann transport, conditional to the probe qubit ${\rm P}$, is performed. As a result, the three qubits $\{\rm S,A,P\}$ are in the superposition
\begin{equation}
|\Phi\rangle_{\rm SAP}=\frac{1}{\sqrt{2}}\big(|\Psi_{\theta(0)}\rangle\otimes|0\rangle_{\rm P}+|\Psi_{\theta(t_{\rm f})}\rangle\otimes|1\rangle_{\rm P}\big).
\label{phi2}
\end{equation}

\begin{figure*}[t]
\includegraphics[width=\textwidth]{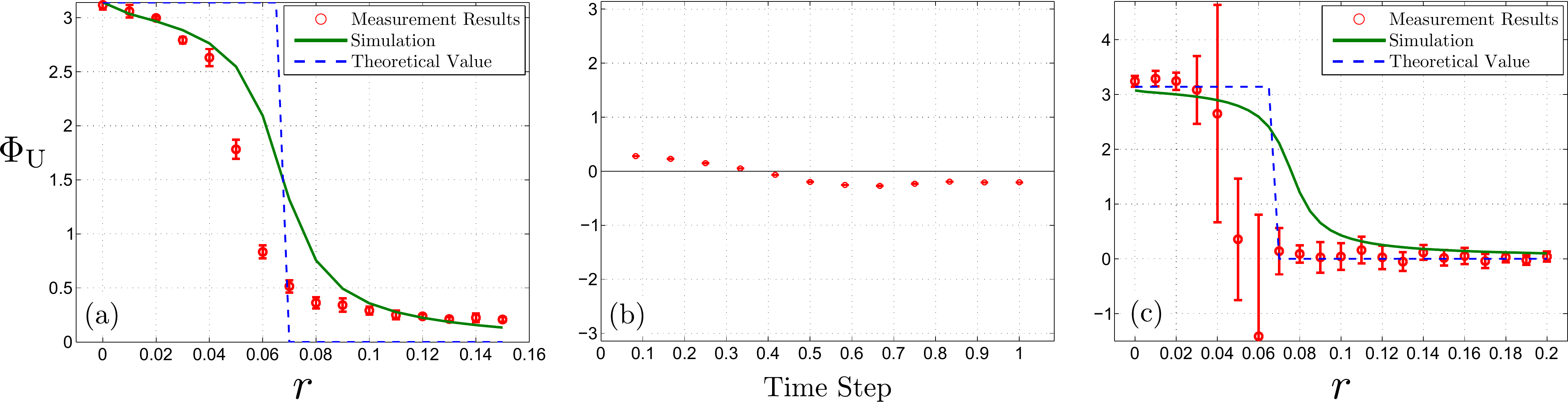}
\caption{Experimental results for the Uhlmann phase $\Phi_{\rm U}$ as a function of the mixedness parameter $r$, defined in Eq.~\eqref{rho1}, or the time step in the holonomy. The red dots with error bars represent the experimental measurements, the dashed blue line is the theoretical value ($r_c\approx0.067$) and the green solid line accounts for a simulation based on an error model (see Methods). In (a), we take $M=0.2$ and measure $\Phi_{\rm U}$ as a function of the mixedness $r$ using $p_a = p_r$, i.e. fulfilling the parallel transport condition. We can clearly see the critical jump from $\Phi_{\rm U}=\pi$ (topological) to $\Phi_{\rm U}=0$ (trivial). In (b), we plot, for $M = 0.2$, $r=0.02$ and $t_{\rm f}=1$, the relative phase $\Phi$ between adjacent states for small time steps $\delta t=0.1$, checking the Uhlmann parallel condition, which implies $\Phi=0$. In (c), we measure $\Phi_{\rm U}$ using the state-independent protocol for $M = 0.6$ and $t_{\rm f}=0.6$, not assuming prior knowledge of the mixedness $r$. The topological transition is clearly appreciable despite the presence of experimental imperfections. The experimental accuracy in $r$ is about $0.01$.}
\label{Fig_noise}
\end{figure*}

\textbf{Step 3.} After the holonomic evolution has been completed, we read out $\Phi_{\rm M}$ from the state of the probe qubit. Tracing out the system and ancilla in Eq. \eqref{phi2}, the reduced state for the probe qubit is
\begin{equation}
\rho_{\rm P}=\frac{1}{2}\Big(\mathds{1}+\text{Re}(\langle\Psi_{\theta(0)}|\Psi_{\theta(t_{\rm f})}\rangle){\bm \sigma_x}+\text{Im}(\langle\Psi_{\theta(0)}|\Psi_{\theta(t_{\rm f})}\rangle){\bm \sigma_y}\Big).
\label{rhop}
\end{equation}
Thus, by measuring the expectation values  $\langle{\bm \sigma_{x}}\rangle$ and $\langle{\bm \sigma_{y}}\rangle$ (green block of Fig.~\ref{Fig_circuit}), we can retrieve $\Phi_{\rm M}$ in the form
\begin{equation}
\begin{split}
\Phi_{\rm M}&=\arg{[\langle{\bm \sigma_{x}}\rangle+{\rm i}\langle{\bm \sigma_{y}}\rangle]}=\\
&=\arg{[\langle\Psi_{\theta(0)}|U_{\rm S}(t_{\rm f})\otimes U_{\rm A}(t_{\rm f})|\Psi_{\theta(0)}\rangle]}.
\end{split}
\label{phase_overlap}
\end{equation}

In Fig.~\ref{Fig_noise} we present the results of phase measurements performed on the IBM Quantum Experience platform \cite{IBMQX}, using three transmon qubits coupled through co-planar waveguide resonators (see Methods). In Fig.~\ref{Fig_noise}(a), we show the measurement of the Uhlmann phase $\Phi_{\rm U}$ for different values of the mixedness parameter $r$, where we set $M=0.2$ and $p_a = p_r$, i.e. fulfilling the parallel transport condition. The critical jump from $\Phi_{\rm U}=\pi$ (topological) to $\Phi_{\rm U}=0$ (trivial) is clearly observed following the previous protocol. 

Additionally, we can check whether the Uhlmann parallel transport condition is satisfied at every time interval during the experiment. By partitioning the closed trajectory in small time steps $\delta t$, the relative phase between the state at time $n\delta t$ and at $(n+1)\delta t$ must be close to zero if the condition is fulfilled. This is the case in the experiment as shown in Fig.~\ref{Fig_noise}(b). During the state preparation (Step 1), we need to include two additional single qubit rotations $R_y^{\alpha_1^{n\delta t}}$ and $R_y^{\alpha_2^{n\delta t}}$ acting on the system and ancilla qubits respectively, where $\alpha_1^{n\delta t}=2I_{0}^{n\delta t}$ and $\alpha_2^{n\delta t}=p_r\alpha_1^{n\delta t}$. These two unitaries make the entangled state between system and ancilla evolve until the state $|\Psi_{n\delta t}\rangle$ is reached. In Step 2, the state evolves to $|\Psi_{(n+1)\delta t}\rangle$ conditional to the state of the probe $P$. The measurement scheme (Step 3) to retrieve the relative phase in Fig.~\ref{Fig_noise}(b) remains the same. Technical details are described in the Supplementary Note 3.
We have included a simulation --green solid line in Fig.~\ref{Fig_noise}-- based on experimental imperfections, mainly finite coherence time ($\sim 50$ $\mu$s) and spurious terms accounting for certain type of electromagnetic crosstalk between qubits. A more detailed description of the error model is given in Methods.

\subsection*{State-independent protocol}

The application of $U_{\rm S}(t)$ and $U_{\rm A}(t)$ with $p_a=p_r$ to the purification $|\Psi_{\theta(t)}\rangle$ implements the Uhlmann parallel transport and hence $\Phi_{\rm M}=\Phi_{\rm U}$. However, this would imply some knowledge about the mixedness parameter $r$ beforehand, which is not always possible. Hence, we present a modification of the previous protocol to measure the topological Uhlmann phase without prior knowledge of the state $\rho$ and its mixedness parameter $r$.

Firstly, we fix $\theta(t)=2\pi t$ and consider open holonomies $\frac{1}{2}<t_{\rm f}<1$ covering more than one half of the complete path. No previous knowledge of the state is assumed to perform the evolution. Hence, the ancillary weight $p_a$ can be different than $p_r$ in Eq.~\eqref{psi2}, but still satisfying $0\le p_a \le 1$. From Eq.~\eqref{phase_overlap2}, the overlap $\langle\Psi_{\theta=0}|\Psi_{\theta=2\pi t_{\rm f}}\rangle$ is always real and thus the phase $\Phi_{\rm M}$ is either $0$ or $\pi$, depending on both the weight $p_r$ associated to the state $\rho_{\theta}$ [Eq.~\eqref{rho1}] and the ancillary weight $p_a$.

We aim to find an $r$-independent value for $p_a$, such that the observed phase $\Phi_{\rm M}$ takes on the same value as the Uhlmann phase for a Hamiltonian with the form of \eqref{H1}. By studying $\Phi_{\rm M}$ as a function of the applied $p_a$, we conclude that if we tune the ancillary weight
\begin{equation}
p_a=p_T:=\frac{-1}{I_{0}^{t_{\rm f}}}\text{arctan}{\Big(\frac{2}{\tan{(I_{0}^{t_{\rm f}})}}\Big)},
\label{paT}
\end{equation}
the value of the observed phase $\Phi_{\rm M}(p_a=p_T)$ coincides with the topological Uhlmann phase $\Phi_{\rm U}$. Algebraic details are provided in Methods. 

Note that there is an intuitive reason why we can get topological information out of a phase associated to a open path longer than one half of a non-trivial topological loop. Indeed, $h(t)$ is symmetric around $t=\frac{1}{2}$. Then, once we have covered one half of the path, we know about the topology of the whole system thanks to this symmetry. Therefore, even an open path for $\frac{1}{2}<t_{\rm f}<1$ can be considered as global.

In terms of the experimental protocol, we only need to modify Step 2 by fixing $p_a=p_T$ for the unitary $U_{\rm A}(t)$. In Fig.~\ref{Fig_noise}(c), we present the results for the state-independent protocol recovering the topological Uhlmann phase without prior knowledge of the state, for $M = 0.6$ and $t_{\rm f} = 0.6$. These are qualitatively the same as in Fig.~\ref{Fig_noise}(a), but the state-independent protocol is more sensitive to errors mainly around the transition point. The mismatch between experiment and simulations is most likely caused by small calibration-dependent systematic errors in the cross-resonance gates.

\section*{DISCUSSION}

We have successfully measured the topological Uhlmann phase, originally proposed in the context of topological insulators and superconductors, making use of ancilla-based protocols. The experiment is realised within a minimal quantum simulator consisting of three superconducting qubits. We have exploited the quantum simulator to realize a controlled coupling of the system to an environment represented by the ancilla degrees of freedom.
Moreover, we have proposed and tested a state-independent protocol that allows us to classify states of topological systems according to the Uhlmann measure. To our knowledge, this is the first time that a noise/temperature induced topological transition in a quantum phase is observed. Recently, these transitions have been addressed in connection to new thermodynamical properties of these systems \cite{Kempkes_et_al_16}. The fact that these effects can be experimentally observed opens the possibility for the search of warm topological matter in the lab. Due to the intrinsic geometric character of the Uhlmann phase, our results may find application in generalisations of holonomic quantum protocols for general, possibly mixed, states.

In addition, an increase of experimental resources such as the number of qubits, the speed and fidelity of the quantum gates, etc. will allow us to study additional topological phenomena with superconducting qubits. In particular, by including interactions in the model Hamiltonian we can test different features: quantum simulations of thermal topological transitions in 2D TIs and TSCs, the interplay between noise and interactions within a topological phase, etc. These effects can be achieved since a system with more interacting qubits can be mapped onto models for interacting fermions with spin \cite{Roushan_et_al14}.
Further details can be found in the Supplementary Note 5. Although such a proposal would be experimentally more demanding, it represents a clear outlook that would need precise controllability of more qubits and the ability to perform more gates with high fidelity.

\subsection*{Data availability}
All relevant data are available from the authors on reasonable request.

\section*{METHODS}

\subsection*{Superconducting Qubit Realization of a Controllable Uhlmann phase.}
\label{app_A}

The experiments on the topological Uhlmann phase have been realized on the IBM Quantum Experience (ibmqx2) \cite{IBMQX}, a quantum computing platform with online user-access based on five fixed-frequency transmon-type qubits coupled via co-planar waveguide resonators.  We have used three qubits, qubit Q0 as the probe qubit, Q1 as the system qubit and Q2 as the ancilla qubit. This choice is motivated by the connectivity required for the measurement protocol and the superior $T_1$ and $T_2$ times of this set of qubits when compared to the set $\{Q2, Q3, Q4\}$ at the time of the experiment. We have used the open-source python SDK \emph{QISKit} \cite{qiskit} to program the quantum computer and retrieve the data. The explicit quantum algorithm to measure the expectation values of $\sigma_x$ and $\sigma_y$ is provided in Supplementary Note 4 using the \emph{OPENQASM} intermediate representation \cite{openqasm}. 
The phase is then extracted from the measured data by evaluating $\Phi_{\rm M} = \arg(\langle\sigma_x\rangle - i\langle \sigma_y\rangle)$. 

For all experiments we have measured 8192 repetitions providing a single value for the phase. For the measurement of the topological Uhlmann phase (Fig.~\ref{Fig_noise}(a)) we vary the initial mixedness of the system state $r$ by setting the rotation angle $\gamma = 2\arccos(\sqrt{1-r})$. The transport of the state according to Uhlmann's parallel transport condition is set by the value $\beta_1 = I_{\rm f}(0,1) = \pi$ for $M<1$ and $\beta_2 = p_a I_{\rm f}(0,1) = 2\pi\sqrt{r(1-r)}$, as defined in Eq.~\eqref{phase_overlap2}. The energy relaxation times of the qubits are $\{T_1^{Q0},\,T_1^{Q1},\,T_1^{Q2}\} = \{45~\mu\rm{s},\,31~\mu\rm{s},\,46~\mu\rm{s}\}$ and the decoherence times $\{T_2^{Q0},\,T_2^{Q1},\,T_2^{Q2}\} = \{40~\mu\rm{s},\,27~\mu\rm{s},\,80~\mu\rm{s}\}$ as stated in the calibration data. 

For the state-independent protocol [Fig.~\ref{Fig_noise}(c), main text] we set $M = 0.6$ and the final time $t_{\rm f} = 0.6$. The system is rotated about $\beta_1 = I_{0}^{0.6} = 2.18537$ and $\beta_2 = p_T I_{0}^{0.6} = 0.954407$. In this measurement energy relaxation and decoherence times are $\{T_1^{Q0},\,T_1^{Q1},\,T_1^{Q2}\} = \{41~\mu\rm{s},\,52~\mu\rm{s},\,62~\mu\rm{s}\}$ and $\{T_2^{Q0},\,T_2^{Q1},\,T_2^{Q2}\} = \{31~\mu\rm{s},\,37~\mu\rm{s},\,87~\mu\rm{s}\}$. Note, that here the error bars are larger as compared to the state-dependent measurement described above, because the expectation values $\langle \sigma_x\rangle$ and $\langle \sigma_y \rangle$ are closer to zero leading to larger statistical errors in the phase. Also, we notice a systematic offset of $\bar{\sigma}_y = 0.098\pm 0.014$ from the expected value $\langle \sigma_y\rangle_{\rm{th}} = 0$. Here, $\bar{\sigma}_y$ is the average over all $r$ values and repetitions. This offset is subtracted from the phase data $\Phi_{\rm M} = \arg[\langle\sigma_x\rangle - i(\langle \sigma_y\rangle-\bar{\sigma}_y)]$ and the result is plotted in Fig.~\ref{Fig_noise}(c). We consider accumulated phase shifts during two-qubit operations as the main reason for this mismatch. We have also noticed that this value changes for different calibrations of the IBM Quantum Experience and when taking different sets of qubits.

Finally, for the measurement of the parallel transport condition we modify the algorithm to prepare the intermediate state $|\Psi_{\theta(n\delta t)}\rangle$ by applying $U_{S/A}(n\delta t)$  to system and ancilla qubit. For the measurement of the Uhlmann phase, the same circuit as above is used to obtain a state evolution $|\Psi_{\theta(n\delta t)}\rangle \rightarrow |\Psi_{\theta((n+1)\delta t)}\rangle$. The complete protocol to measure the parallel transport condition is shown in the Supplementary Fig. 1. In the experiment, we choose $M=0.2$ and $r=0.02$ to stay within the topological sector. The mixedness angle evaluates to $\gamma = 2\arccos(\sqrt{0.95}) = 0.2838$. The angles for the intermediate state preparation are determined by $\alpha_1(n) = I_{0}^{n\delta t}$ and $\alpha_2(n) = p_r I_{0}^{n\delta t}= 2 \sqrt{r(1 - r)} I_{0}^{n\delta t} = 0.28 I_{0}^{n\delta t}$, 
the evolution from $n\delta t$ to $(n+1)\delta t$ is determined by the angles $\beta_1(n) = I_{n\delta t}^{(n+1)\delta t}$ and $\beta_2(n) = p_r I_{n\delta t}^{(n+1)\delta t} = 0.28 I_{n\delta t}^{(n+1)\delta t}$. The recorded data shown in Fig. 3(b), main text, shows that the measured phase difference $\langle \Phi_{\rm M}(n\delta t) \rangle = -0.07 \pm 0.2$ is zero within the statistics. However, the residuals do not follow a normal distribution which hints at systematic gate errors instead of stochastic errors.

\subsection*{State-independent Derivation}
\label{app_B}

The derivation of the value for $p_T$ [Eq.~\eqref{paT}] is as follows.
From Eq.~\eqref{phase_overlap2} we find the value $p_a=p^{c}_a$ (where the superindex \emph{c} stands for critical) at which $\Phi_{\rm M}$ goes abruptly from $\pi$ to $0$ as a function of $p_r$ and $I_{\rm f}$,
\begin{equation}
p_a^c=\frac{-1}{I_{0}^{t_{\rm f}}}\arctan{\Big(\frac{1}{p_r\tan{(I_{0}^{t_{\rm f}})}}\Big)}.
\label{pac}
\end{equation}
If we set $\frac{1}{2}<t_{\rm f}<1$, then $p_a^c$ is a monotonically decreasing function of $p_r$,
\begin{equation}
\frac{\partial p_a^c}{\partial p_r}=\frac{\tan{(I_{0}^{t_{\rm f}})}}{I_{0}^{t_{\rm f}}[1 + p_r^2\tan^2{(I_{0}^{t_{\rm f}})}]}<0.
\label{Dpac}
\end{equation}
If $M>1$, then $-\pi/2<I_{0}^{t_{\rm f}}<\pi/2$, which from Eq.~\eqref{phase_overlap2} implies that $\Phi_{\rm M}=0$ for any value of $p_r$ and $p_a$. Hence, for the trivial case $M>1$, there is no critical value $p_a^c$ and $\Phi_{\rm M}=0$ always. This maps $\Phi_{\rm M}$ to the Uhlmann phase $\Phi_{\rm U}$ at least for this case.
On the contrary, if $M<1$, then $\pi/2<I_{0}^{t_{\rm f}}<\pi$ which implies $\tan{(I_{0}^{t_{\rm f}})}<0$. Since $0<p_r<1$, then $-\arctan{\Big(\frac{1}{p_r\tan{(I_{0}^{t_{\rm f}})}}\Big)}<\pi/2$. Thus, there is always a solution of Eq.~\eqref{pac} with $0<p_a^c<1$ for any $p_r$.
As discussed in the main text, the state $\rho_\theta$ in Eq.~\eqref{rho1} is topological in the Uhlmann sense $\Phi_{\rm U}=\pi$, only if $M<1$ and $p_r<0.5$.

Now, we define $p_{\rm T}:=p_a^c(p_r=0.5)$ using Eq.~\eqref{pac}. Note that the true $p_r$ of the system is unknown as we have assumed no knowledge of the state. Nevertheless, if $p_r>0.5$, then its associated critical value [from Eq.~\eqref{pac}] is $p_a^c<p_{\rm T}$. This means that by applying $U_{\rm A}$ with $p_a=p_{\rm T}$ and measuring the associated phase $\Phi_{\rm M}$ we can extract the following conclusions:
\begin{itemize}
\item If we measure $\Phi_{\rm M}(p_{\rm T})=0$, the system is within a trivial phase ($\Phi_{\rm U}=0$). Because this implies $p_a^c<p_{\rm T}$ and hence $p_r>0.5$ ($\Phi_{\rm U}=0$), as we have proven that $p_a^c$ always decreases with $p_r$.
\item If we measure $\Phi_{\rm M}(p_{\rm T})=\pi$, the system is in a topological phase ($\Phi_{\rm U}=\pi$). Because in that case $p_a^c>p_{\rm T}$ and then $p_r<0.5$ ($\Phi_{\rm U}=\pi$).
\end{itemize}
Hence, we have just proven that $\Phi_{\rm M}(p_{\rm T})=\Phi_{\rm U}$. \\

\subsection*{Error Simulation}
\label{app_C}

The detrimental effect of experimental errors is modelled by means of a Liouvillian term $\mathcal{L}_{\rm error}$, so that the Liouvillian $\mathcal{L}_0$, accounting for the idealized dynamics, is in fact substituted by $\mathcal{L}_0+\mathcal{L}_{\rm error}$. Specifically, if a gate is performed during a time $\tau$ via a Hamiltonian $H_0$, i.e. $U_{\rm gate}=e^{-iH_0\tau}$, we substitute
\begin{equation}
e^{-iH_0\tau}\rho e^{iH_0\tau}\equiv e^{\mathcal{L}_0\tau}\rho\rightarrow e^{(\mathcal{L}_0+\mathcal{L}_{\rm error})\tau}\rho.
\end{equation}
This error Liouvillian includes typical sources of imperfections: a) a residual IX term during the cross-resonance ZX90 gate in the implementation of the CNOTs, $H_{\rm ZX}=mIX+\mu ZX$ \cite{ZX90_1,ZX90_2,ZX90_3}; b) spontaneous emission and dephasing terms $\mathcal{L}_-(\rho)=\gamma_-(\sigma_-\rho\sigma_+-\tfrac{1}{2}\{\sigma_+\sigma_-,\rho\})$ and $\mathcal{L}_z(\rho)=\gamma_z(\bm{\sigma}_z\rho\bm{\sigma}_z-\rho)$, respectively.

We have accommodated the values of $\gamma_-$ and $\gamma_z$ to the characteristic longitudinal and transverse relaxation times of $T_1=51\,\mu \text{s}$ and $T_2=51\,\mu \text{s}$ reported by the IBM Quantum Experience calibration team the day of the measurements. The residual IX strength has been taken to be about $m\sim 0.4~ \text{MHz}$. In addition, we consider $\tau_{2\pi}\sim 200~\text{ns}$ and $\tau_{\rm ZX90}\sim600~\text{ns}$ as characteristic times for a $2\pi$-rotation on a single qubit and the ZX90 gates, respectively. Waiting times of 5 ns after a single qubit gate and 40 ns after a ZX90 gate are also included.  

In Fig.~\ref{Fig_noise}, we plot  the result of the simulation including these experimental imperfections together with the experimental measurements of the topological Uhlmann phase $\Phi_{\rm U}$. Despite the errors, the topological transition is clearly noticed.

\section*{Acknowledgments}

M.A.MD., A.R. and O.V. thank the Spanish MINECO grant FIS2012-33152, a ``Juan de la Cierva-Incorporaci\'on'' reseach contract, the CAM research consortium QUITEMAD+ S2013/ICE-2801, the U.S. Army Research Office through grant W911NF-14-1-0103, Fundaci\'on Rafael del Pino, Fundaci\'on Ram\'on Areces, and RCC Harvard. S.G., A.W. and S.F. acknowledge support by the Swiss National Science Foundation (SNF, Project 150046).

\section{SUPPLEMENTARY NOTES}

\appendix

\setcounter{figure}{0}
\setcounter{equation}{0}
\renewcommand*{\thefigure}{S\arabic{figure}}
\renewcommand*{\theequation}{S\arabic{equation}}

\subsection{Supplementary Note 1: Uhlmann phase for qubit systems}
\label{app_A}

The Uhlmann phase extends the notion of the geometric Berry phase from pure quantum states (Berry) to mixed quantum states described by density matrices. Uhlmann was first to study this problem from a rigorous mathematical perspective \cite{Uhlmann86} and to find a satisfactory solution \cite{Uhlmann89, Uhlmann91, Hubner93, Uhlmann96}.

Let $\theta(t)|_{t=0}^1$ define a trajectory along a family of single qubit density matrices parametrised by $\theta$,
\begin{equation}\label{rho1}
\rho_\theta=(1-r)|0_{\theta}\rangle\langle 0_{\theta}| + r|1_{\theta}\rangle\langle 1_{\theta}|,
\end{equation}
where $r$ stands for the degree of mixedness between the ground state $|0_{\theta}\rangle$ and the excited state $|1_{\theta}\rangle$. Note that $\rho_{\theta}$ can always be viewed as a pure state $|\Psi_{\theta}\rangle$ in an enlarged Hilbert space ${\cal H}={\cal H}_{\rm S}\otimes{\cal H}_{\rm A}$, where $S$ stands for system and $A$ for the ancilla degrees of freedom. This process is called purification, and satisfies the constraint $\rho_{\theta}={\rm Tr_A}\Big(|\Psi_{\theta}\rangle\langle\Psi_{\theta}|\Big)$. The set of purifications $|\Psi_{\theta}\rangle$ generates the family of density matrices $\rho_{\theta}$. This aims to be the density-matrix analog to the standard situation where vector states $\ket{\psi}$ span a Hilbert space and generate pure states by the relation $\rho=|\psi\rangle\langle\psi|$.
Actually, the phase freedom of pure states, U(1)-gauge freedom, is generalised to a ${\rm U}(n)$-gauge freedom ($n$ is the dimension of the density matrix). This occurs since $|\Psi_{\theta}\rangle$ and $V^{\rm t}_{\rm A}|\Psi_{\theta}\rangle$  are purifications of the same density matrix for some unitary operator $V^{\rm t}_{\rm A}$ acting on the ancilla degrees of freedom. The superindex ${\rm t}$ denotes the transposition with respect to the qubit eigenbasis.
If the trajectory defined by $\theta(t)$ is closed $\rho_{\theta(1)}=\rho_{\theta(0)}$, the initial and final purifications must differ only in a unitary transformation $V^{\rm t}_{\rm A}$, $|\Psi_{\theta(1)}\rangle=V^{\rm t}_{\rm A}|\Psi_{\theta(0)}\rangle$. Hence, by analogy to the pure state case, Uhlmann defines a parallel transport condition such that $V_{\rm A}$ is constructed by imposing that the distance between two infinitesimally closed purifications, $\|\Psi_{\theta(t+dt)}\rangle-|\Psi_{\theta(t)}\rangle\|^2$, reaches its minimum value. Then it is possible to write

\begin{equation}
V_{\rm A}=\mathcal{P}{\rm e}^{\int A_{\rm U}},
\label{VA}
\end{equation}
where $\mathcal{P}$ stands for the path ordering operator along the trajectory $\theta(t)|_{t=0}^1$, and $A_{\rm U}$ is the so-called Uhlmann connection form \cite{Uhlmann86,Viyuela_et_al14,Viyuela_et_al15}.

The Uhlmann geometric phase is defined from the mismatch between the initial point $|\Psi_{\theta(0)}\rangle$ and the final point after parallel transport, i.e. $|\Psi_{\theta(1)}\rangle$,

\begin{equation}\label{phiU1}
\Phi_{\rm U}:=\arg{[\langle\Psi_{\theta=0}|\Psi_{\theta=1}\rangle]}=\arg \Tr\left[\rho_{\theta(0)}V_{\rm A}\right].
\end{equation}
This phase is a gauge independent quantity \cite{Uhlmann86,Uhlmann89,Uhlmann91}, that comes from the parallel transport of the purification $|\Psi_{\theta}\rangle$. The most explicit formula for the Uhlmann connection was given by H\"ubner \cite{Hubner93},
\begin{equation}\label{A1}
A_{\rm U}=\sum_{i,j}|\psi^i_{\theta}\rangle\frac{\langle\psi^i_{\theta}|\left[(\partial_{\theta}\sqrt{\rho_{\theta}}),\sqrt{\rho_{\theta}}\right]|\psi^j_{\theta}\rangle}{p^i_{\theta}+p^j_{\theta}}\langle\psi^j_{\theta}|d\theta,
\end{equation}
in the spectral basis of $\rho_{\theta}=\sum_jp^j_{\theta}|\psi^j_{\theta}\rangle\langle\psi^j_{\theta}|$. The parameter $\theta$ may play the role of the crystalline momentum in condensed matter systems.

The derivative of the square-root of the density matrix with respect to the transport parameter $\theta$ is given by

\begin{eqnarray}\label{drho1}
\partial_{\theta}\sqrt{\rho_\theta}&=&\sqrt{(1-r)}\big(|\partial_{\theta}0_{\theta}\rangle\langle 0_{\theta}|+|0_{\theta}\rangle\langle\partial_{\theta} 0_{\theta}|\big)+\nonumber\\
&+&\sqrt{r}\big(|\partial_{\theta}1_{\theta}\rangle\langle 1_{\theta}|+|1_{\theta}\rangle\langle\partial_{\theta}1_{\theta}|\big).
\end{eqnarray}

We can simplify the connection $A_{\rm U}$ in Eq.~\eqref{A1}, for the density matrix \eqref{rho1} and the Hamiltonian (3) in the main text. We substitute Eq.~\eqref{drho1} in Eq.~\eqref{A1}, and take into account that the summation indices in Eq.~\eqref{A1} only runs over the states $|1_{\theta}\rangle$ and $|0_{\theta}\rangle$, obtaining 
\begin{align}
A_{\rm U}=\Big[(1&-p_r)\langle 1|\partial_{\theta}0_{\theta}\rangle~|0_{\theta}\rangle\langle 1_{\theta}| \nonumber\\
&+ (1-p_r)\langle 0_{\theta}|\partial_{\theta}1_{\theta}\rangle~|0_{\theta}\rangle\langle 1_{\theta}|\Big]d\theta,
\label{A2}
\end{align}
where $p_{r}=2\sqrt{r(1-r)}$. \\

For computational purposes, we fix the gauge for the eigenstates of the system Hamiltonian in Eq.~(3) such that, 
\begin{eqnarray}
|0_{\theta}\rangle&=&\frac{1}{\sqrt{1+g^2(\theta,M)}} \begin{pmatrix} 1 \\ g(\theta,M) \end{pmatrix}, \label{eigen1}\\
|1_{\theta}\rangle&=&  \frac{1}{\sqrt{1+g^2(\theta,M)}} \begin{pmatrix} g(\theta,M) \\ -1 \end{pmatrix},
\label{eigen2}
\end{eqnarray}
where
\begin{equation}
g(\theta,M):=\frac{\sin{\theta}}{M + \cos{\theta}+\sqrt{1 + M^2 + 2M\cos{\theta}}}.
\label{gt}
\end{equation}

From Eq.~\eqref{eigen1} and Eq.~\eqref{eigen2}, we compute
\begin{eqnarray}
\langle0_{\theta}|\partial_{\theta}1_{\theta}\rangle=\frac{\partial_{\theta}{n}_{\theta}^x}{2n_{\theta}^z}&=&\frac{1+M\cos{\theta}}{2+2M^2+4M\cos{\theta}},\nonumber\\
|0_{\theta}\rangle\langle 1_{\theta}|-|1_{\theta}\rangle\langle 0_{\theta}|&=&\begin{pmatrix} 0 & -1 \\ 1 & 0 \end{pmatrix},
\label{overlaps}
\end{eqnarray}
where ${n}_{\theta}^i$ is the $i-$th component of the winding vector. Finally, we insert Eqs.~\eqref{overlaps} in Eq.~\eqref{A2} and obtain
\begin{equation}
A_{\rm U}=-{\rm i}(1-p_r)\frac{\partial_{\theta}{n}_{\theta}^x}{2n_{\theta}^z}{\bm \sigma_y}d\theta.
\label{A3}
\end{equation}

As the connection in Eq.~\eqref{A3} commutes for different values of $\theta$, we can drop the path ordering that appears in the expression for the Uhlmann unitary [Eq.~\eqref{VA}], and get the simplified equation
\begin{equation}
V_{\rm A}({\theta})={\rm e}^{-{\rm i}(1-p_r)\int_0^{\theta}\frac{\partial_{\theta '}{n}_{\theta '}^x}{2n_{\theta '}^z}{\bm \sigma_y}d\theta '}.
\label{VA2}
\end{equation}

Lastly, we substitute Eq.~\eqref{VA2} and Eq.~\eqref{rho1} in Eq.~\eqref{phiU1} to compute the Uhlmann phase
\begin{equation}
\Phi_{\rm U}=\arg\Bigg\{\cos\bigg[\frac{1-2p_r}{2}\int_0^{\theta}\bigg(\frac{\partial_{\theta'} {n}_{\theta'}^x}{n_{\theta'}^z}\bigg)d{\theta'}\bigg]\Bigg\}.
\label{phiU2b}
\end{equation}

The mapping $\rho \longrightarrow V_{\rm A}$ is a so-called pointed holonomy. This means, that even if the trajectory in parameter space is closed, in general the holonomy depends on the initial point of the path \cite{Viyuela_et_al15,Budich_et_al15}. Nonetheless, we have identified instances in which the pointed holonomy reduces to an \emph{absolute holonomy} becoming independent of the initial point \cite{Viyuela_et_al14,Viyuela_et_al14_2D}. This is indeed the case studied in the present paper, as well as most of the representative models of 1D and 2D topological insulators and superconductors.

\subsection{Supplementary Note 2: Holonomic time evolution}
\label{app_B}

At this stage, we would like to physically implement the holonomy that has been mathematically described in the previous section. For that purpose, we express the parallel transport generated by the change in the parameter $\theta$, as a unitary time evolution over system and ancilla $U_{\rm S}\otimes U_{\rm A}$ where the control-parameter $\theta(t)$ is varied in time. The system unitary evolution $U_{\rm S}$ is defined through the relations
\begin{equation}
|0\rangle_{\theta(t)}:=U_{\rm S}(t)|0\rangle,~~|1\rangle_{\theta(t)}:=U_{\rm S}(t)|1\rangle,
\end{equation}
where $\ket{0}=\begin{pmatrix}1 \\ 0 \end{pmatrix}$ and $\ket{1}=\begin{pmatrix}0 \\ 1 \end{pmatrix}$ is the standard qubit basis. Using the eigenstate equations \eqref{eigen1} and \eqref{eigen2}, $U_{\rm S}(\theta)$ is obtained straightforwardly,
\begin{equation}
U_{\rm S}(t)=\frac{1}{\sqrt{1+g^2[\theta(t),M]}} \begin{pmatrix} 1 & - g[\theta(t),M] \\  g[\theta(t),M] & 1 \end{pmatrix},
\label{Utheta1}
\end{equation}
where $g(\theta,M)$ was defined in Eq.~\eqref{gt}.

At this point Eq.~\eqref{Utheta1} can be expressed as the exponential of a Hamiltonian using the following relations
\begin{eqnarray}
U_{\rm S}(t)&=&{\rm e}^{-{\rm i}\int_0^{t}h(t')dt'},\label{Ut1b}\\
h(t)&=&{\rm i}\bigg(\frac{d\theta}{dt}\bigg)\big[\partial_{\theta}U_{\rm S}(\theta)\big]U^{\dagger}_{\rm S}(\theta).
\label{Uformula}
\end{eqnarray}

We substitute Eq.~\eqref{Utheta1} into Eq.~\eqref{Uformula}, arriving at
\begin{equation}
h(t)=\bigg(\frac{d\theta}{dt}\bigg)\frac{\partial_{\theta(t)}{n}_{\theta(t)}^x}{2n_{\theta(t)}^z}{\bm \sigma_y},
\label{Htb}
\end{equation}
where we have used Eq.~\eqref{overlaps} as well.

The unitary for the ancilla qubit $U_{\rm A}$ is determined by combining: 1) the transport of the eigenstates $\ket{0(1)_\theta}$ through $U_{\rm S}(t)$ and 2) the Uhlmann correction $V_{\rm A}[\theta(t)]$, for the purification as a whole to be parallely transported [Eq.~\eqref{VA2}]; hence,
\begin{equation}
U_{\rm A}(t)=[U_{\rm S}^\dagger(t) V_U(t)]^{\rm t}.
\label{UA}
\end{equation}
Here, the superindex ${\rm t}$ denotes the transposition with respect to the qubit eigenbasis. Further simplifications of Eq.~\eqref{UA} using Eq.~\eqref{Htb} and Eq.~\eqref{VA2} lead to
\begin{equation}
U_{\rm A}(t)={\rm e}^{-{\rm i}p_a\int_0^{t}h(t')dt'},
\label{UA2}
\end{equation}
with $p_a=p_r$.


\begin{figure}[t]
\includegraphics[width=\columnwidth]{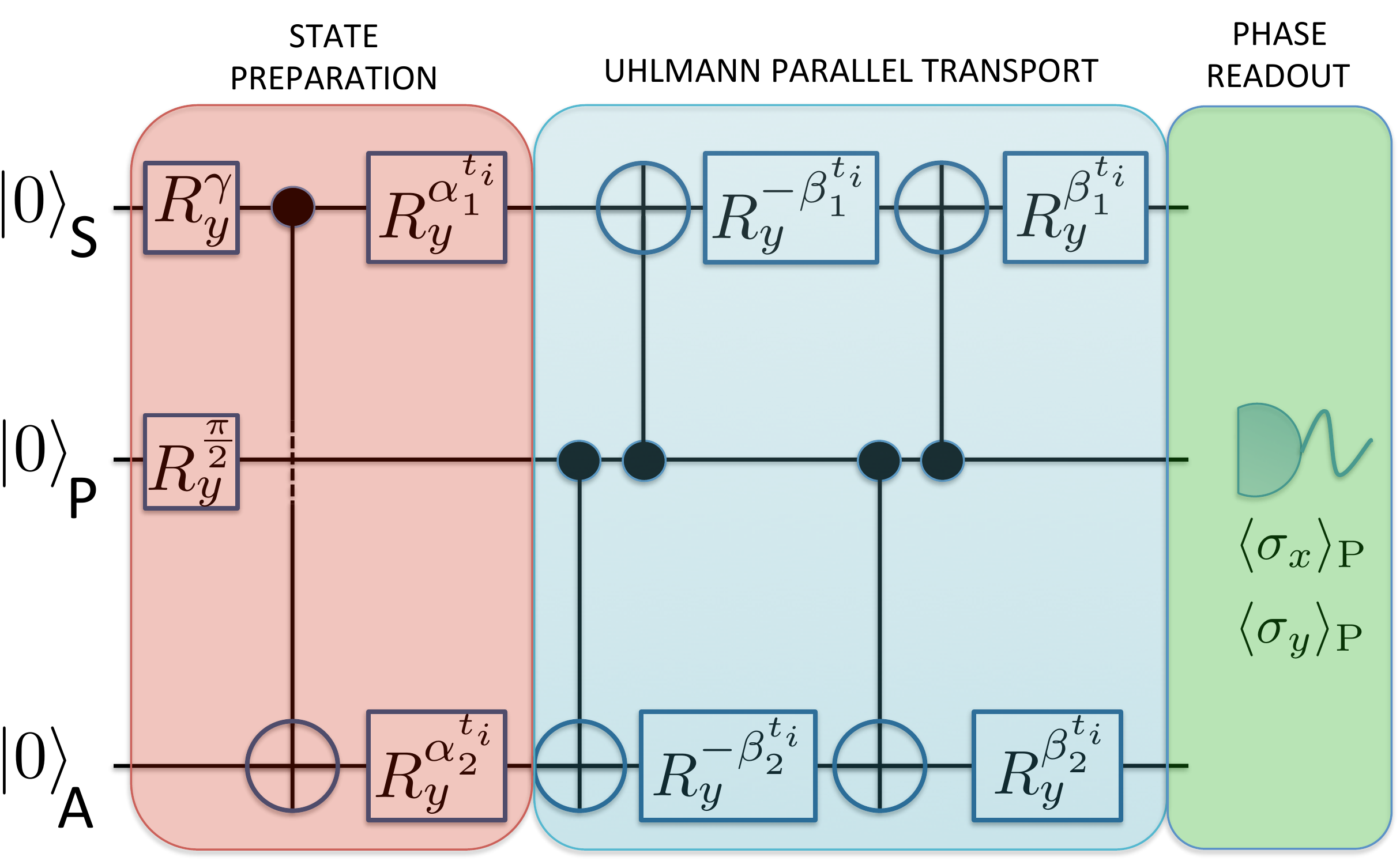}
\caption{\label{protocol3} Circuit diagram to test the Uhlmann parallel transport condition. The gates $R_y^\alpha$ represent single qubit rotations about the y-axis for an angle $\alpha$, and the different angles of rotation are given in Eq.~\eqref{alpha} and Eq.~\eqref{beta}.}
\end{figure}

\subsection{Supplementary Note 3: Experimental test of Uhlmann parallel transport}
\label{app_D}

In this section we present further details on how to experimentally test the Uhlmann parallel transport condition along the holonomy. The experimental results are shown in the middle plot of Fig. 3 in the main text.

The protocol is depicted in Fig.~\ref{protocol3}. The state preparation part of the protocol is the same as in the state-dependent and state-independent protocols described in the main text. We prepare the initial state $|\Psi_{t=0}\rangle$ using single qubit rotations $R_y^\theta$ about the y-axis for an angle $\theta$ and a two-qubit controlled not gate. Then, we apply $R_y^{\alpha_1^{t_i}}$ and $R_y^{\alpha_2^{t_i}}$ on the system and ancilla qubits respectively, where 
\begin{equation}
\begin{split}
\alpha_1^{t_{i}}&=2\int_0^{t_i}h(t')dt',\\
\alpha_2^{t_{i}}&=p_r\alpha_1^{t_{i}}.
\end{split}
\label{alpha}
\end{equation}
The entangled state between system and ancilla will evolve until $|\Psi_{t=t_i}\rangle$ (red block of Fig.~\ref{protocol3}). Next, we apply $R_y^{\beta^{t_{i}}_{1(2)}}$ to the system and ancilla qubits, conditional on the probe (which has been previously prepared on an equal superposition of states $\ket{0}$ and $\ket{1}$). The angles of rotation in this case are
\begin{equation}
\begin{split}
\beta_1^{t_{i}}&=\int_{t_i}^{t_i+\delta t}h(t')dt',\\
\beta_2^{t_{i}}&=p_r\beta_1^{t_{i}}.
\end{split}
\label{beta}
\end{equation}
This part comprises the blue block in Fig.~\ref{protocol3}. As a consequence, the three qubits $\{\rm S,A,P\}$ end up in the superposition
\begin{equation}
|\Phi\rangle_{\rm SAP}=\frac{1}{\sqrt{2}}\big(|\Psi_{t=t_i}\rangle\otimes|0\rangle_P+|\Psi_{t=t_i+\delta t}\rangle\otimes|1\rangle_P\big).
\label{phi2}
\end{equation}
Now we are interested in reading out the relative phase $\Phi$ from the state of the probe qubit. Tracing out the system and ancilla in Eq. \eqref{phi2}, the reduced state for the probe qubit is
\begin{equation}
\begin{split}
\rho_{\rm P}&=\frac{1}{2}\Big(\mathds{1}+\text{Re}(\langle\Psi_{t=t_i}|\Psi_{t=t_i+\delta t}\rangle){\bm \sigma_x}\\
&+\text{Im}(\langle\Psi_{t=t_{i}}|\Psi_{t=t_i+\delta t}\rangle){\bm \sigma_y}\Big).
\end{split}
\label{rhop}
\end{equation}
Thus, by measuring the expectation values $\langle{\bm \sigma_{x}}\rangle$ and $\langle{\bm \sigma_{y}}\rangle$ (green block of Fig.~\ref{protocol3}), we can retrieve the relative phase $\Phi$ between the states $|\Psi_{t=t_i}\rangle$ and $|\Psi_{t=t_i+\delta t}\rangle$. If the transport fulfills the Uhlmann parallel condition between $t_i$ and $t_i + \delta t$, then the two vectors should be in phase $\Phi\approx0$. This is actually what we observe in the experiment (see the middle plot in Fig. 3 of the main text).

\subsection{Supplementary Note 4: Quantum Algorithm}
\label{app_E}

\begin{figure*}[t]
\centering
\includegraphics[width=\textwidth]{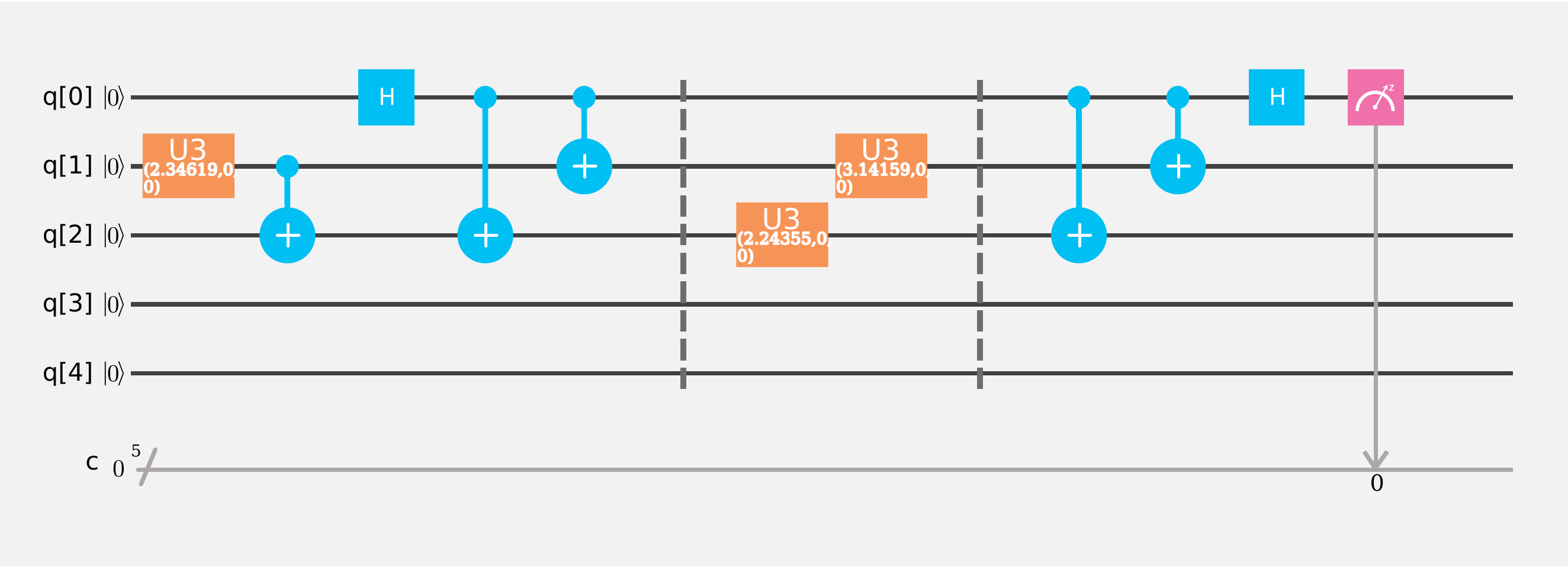}
\caption{\label{fig:statedepr15} Example of the circuit diagram on the IBM Quantum Experience for the measurement of the topological Uhlmann phase for $r=0.15$. The values for the rotation angles are $\gamma = 2\arccos(\sqrt{1-r})=0.7954$, $\beta_1 = \pi$ and $\beta_2 = 2\pi\sqrt{r(1-r)} = 2.24355$}
\end{figure*}

The experiments on the topological Uhlmann phase have been realized on the IBM Quantum Experience (ibmqx2) \cite{IBMQX}, a quantum computing platform with online user-access based on five fixed-frequency transmon-type qubits coupled via co-planar waveguide resonators.  We have used three qubits, qubit Q0 as the probe qubit, Q1 as the system qubit and Q2 as the ancilla qubit. This choice is motivated by the connectivity required for the measurement protocol and the superior $T_1$ and $T_2$ times of this set of qubits when compared to the set $\{Q2, Q3, Q4\}$ at the time of the experiment. We have used the open-source python library \emph{QISKit} \cite{qiskit} to program the quantum computer and retrieve the data. The quantum algorithm to measure the expectation values of $\sigma_x$ is 
\begin{lstlisting}[ mathescape,
  columns=fullflexible,
  basicstyle=\fontfamily{lmvtt}\selectfont,
]
OPENQASM 2.0;
include "qelib1.inc";

qreg q[5];
creg c[5];
u3($\theta$,0,0) q[1];
cx q[1],q[2];
h q[0];
cx q[0],q[2];
cx q[0],q[1];
barrier q[0],q[1],q[2],q[3],q[4];
u3($\beta_2$,0,0) q[2];
u3($\beta_1$,0,0) q[1];
barrier q[0],q[1],q[2],q[3],q[4];
cx q[0],q[2];
cx q[0],q[1];
h q[0]; 
measure q[0] -> c[0];
\end{lstlisting}

\noindent in the \emph{OPENQASM} intermediate representation, described in \cite{openqasm}. For measuring $\sigma_y$ we insert the rotation 'sdg q[0];' before the last hadamard gate 'h [q0]'. The circuit diagram (including the rotation for the measurement of $\sigma_y$) is shown in Figure \ref{fig:statedepr15}. Note, that we do not implement the last set of single-qubit rotations on system and ancilla qubits because these do not change the outcome of the measurement of the probe qubit $Q_0$. 
The phase is then extracted from the measured data by evaluating $\Phi_{\rm M} = \arg(\langle\sigma_x\rangle - i\langle \sigma_y\rangle)$. 

\subsection{Supplementary Note 5: Interacting Systems $\&$ 2D}  
\label{app_G}

The protocol to measure the topological Uhlmann phase deals with single-qubit Hamiltonians [Eq.~(3) in the main text]. These can be mapped to free-fermion topological insulators. We may identify the ramp parameter $\theta$ with the crystalline momentum in the Brillouin zone $k$. In 2D, a way to define a non-trivial topological invariant for isotropic systems at finite temperature is by means of the winding number of the Uhmann phase, as shown in Refs.~\cite{Viyuela_et_al14_2D,Arovas14,Viyuela_et_al15}. We could test this experimentally by mapping two independent parameters $\theta$ and $\delta$ of our quantum simulator to the crystalline momentum of a 2D topological insulator $k_x$ and $k_y$. By measuring the Uhlmann phase along $\theta$, for different values of $\delta$, we can extract the value of the winding number by observing discontinuous jumps in the Uhlmann phase.

More complicated Hamiltonians involving more qubits could be considered in a more general setup. Actually, it has been shown in Ref.~\cite{Roushan_et_al14} that an $L-$qubit interacting system can be mapped onto two types of systems that we discuss in what follows.

On the one hand, a system of 2 qubits can be mapped to a system of two interacting fermions with spin $1/2$. Therefore, an Uhlmann experiment for interacting 2-qubit Hamiltonians would be the first experimental measurement of a topological phase associated to an interacting system in a mixed state. It would be very interesting to analyse how the interaction term counteracts or enhances the effect that noise produces in the system.

On the other hand, there is a complementary mapping from a many-body interacting spin system to Haldane-like models \cite{Haldane_88} with $2^L$ bands. These are free-fermion models but the fact of having more bands opens the possibility of having higher topological quantum numbers. From the point of view of the Uhlmann theory of symmetry-protected topological order at finite temperature, one can envision the possibility of testing topological transitions between non-trivial topological phases solely driven by noise or temperature. This is an effect that only appears in systems with high topological numbers as shown in \cite{Viyuela_et_al14_2D}.

\end{document}